\documentclass[11pt,letterpaper]{amsart}
\usepackage{amssymb}
\usepackage[all]{xy} 

\newtheorem{conjecture}{Conjecture}
\newtheorem{definition}{Definition}
\newtheorem{proposition}{Proposition}
\newtheorem{remark}{Remark}
\newtheorem{example}{Example}
\newtheorem{problem}{Problem}
\newtheorem{theorem}{Theorem}
\theoremstyle{definition}
\newtheorem{open}{Open Problem}
\newcommand{\CF}{\operatorname{CF}}

 \newcommand{\ind}[1]{ } 

\DeclareMathOperator{\GL}{GL}
\DeclareMathOperator{\End}{End}
\DeclareMathOperator{\Sym}{Sym}
\DeclareMathOperator{\perm}{perm}
\DeclareMathOperator{\pdet}{PDet}
\DeclareMathOperator{\mult}{mult} 
\DeclareMathOperator{\imm}{im} 
\DeclareMathOperator{\tr}{tr} 
\newcommand{\Det}{\ensuremath{\mathcal{D}et}} 
\newcommand{\Perm}{\ensuremath{\mathcal{P}erm}} 
\newcommand{\C}{\mathbb{C}}
\newcommand{\R}{\mathbb{R}}
\newcommand{\Z}{\mathbb{Z}}
\newcommand{\N}{\mathbb{N}}
\newcommand{\PP}{\mathbb{P}}
\newcommand{\cc}[1]{\mathsf{#1}} 

\newcommand{\BR}{\mathbb{R}}
\newcommand{\BC}{\mathbb{C}}
\newcommand{\BQ}{\mathbb{Q}}
\newcommand{\BZ}{\mathbb{Z}}
\newcommand{\BF}{\mathbb{F}}
\newcommand{\cPH}{\cc{PH}}
\newcommand{\cHN}{\cc{HN}}
\newcommand{\size}{\operatorname{size}}
\newcommand{\poly}{\operatorname{poly}}
\newcommand{\FEAS}{\operatorname{FEAS}}

\newcommand{\forkorben}[1]{ }

\begin{document}

\title{\mbox{}\\ 
\vspace{-1in}Report on \\
``Mathematical Aspects of $\cc{P}$ vs.\ $\cc{NP}$ and its Variants'' \\
August 1-5, 2011, Institute for Computational and Experimental Research in 
Mathematics (ICERM), Providence, Rhode Island \\
Organizers: Saugata Basu, J. M. Landsberg, J. Maurice Rojas
}
\author{Joshua A.\ Grochow and Korben Rusek}

\begin{abstract}
This is a report on a workshop held August 1 to August 5, 2011 at the Institute for Computational and Experimental Research in Mathematics (ICERM) at Brown University, Providence, Rhode Island, organized by Saugata Basu, Joseph M. Landsberg, and J. Maurice Rojas. We provide overviews of the more recent results presented at the workshop, including some works-in-progress as well as tentative and intriguing ideas for new directions. The main themes we discuss are representation theory and geometry in the Mulmuley--Sohoni Geometric Complexity Theory Program, and number theory and other ideas in the Blum--Shub--Smale model.
\end{abstract}

\maketitle

\pagestyle{myheadings}
\markboth{Report on ``Mathematical Aspects of $\cc{P}$ vs.\ $\cc{NP}$ and its Variants''}{Report on ``Mathematical Aspects of $\cc{P}$ vs.\ $\cc{NP}$ and its Variants''}

\section{Introduction} 
The $\cc{P}$ vs.\ $\cc{NP}$ problem, which can be traced back to a 1956 
letter of G\"odel to von Neumann (and to a recently declassified 1955 letter from Nash to the U. S. National Security Agency), lies at the heart of theoretical computer 
science. This problem underlies not only the computational complexity of 
numerous practical problems: it has deep connections with fundamental 
mathematical questions coming from geometry, representation theory, and number 
theory.  This workshop focuses on recently discovered connections along these 
lines. The workshop also included several presentations on other interesting aspects of complexity such as holographic algorithms, dichotomy theorems, quantum algorithms, and number-theoretic problems.

The new field of Geometric Complexity Theory establishes a 
representation-theoretic approach to a close cousin of the 
$\cc{P}$ vs.\ $\cc{NP}$ problem: the $\cc{VP}$ vs.\ $\cc{VNP}$ problem. 
Moreover, independent of the resolution of the latter question, Geometric 
Complexity Theory leads to beautiful open questions in geometry and 
representation theory. We first review the Geometric Complexity Theory aspect of 
our workshop. Later on, we detail other variants of the 
$\cc{P}$ vs.\ $\cc{NP}$ problem, as well as an approach to the 
original $\cc{P}$ vs.\ $\cc{NP}$ problem coming from number theory. 

\section{Geometric Complexity Theory}
In \cite{gct1,gct2} Mulmuley and Sohoni introduced the Geometric Complexity Theory (GCT) program to approach fundamental problems in complexity theory such as $\cc{P}$ vs.\ $\cc{NP}$. Several active researchers in GCT and the related mathematics presented talks at this workshop.

The permanent versus determinant conjecture is a long-standing conjecture in complexity theory based on the work of Valiant \cite{valiant}:

\begin{conjecture}[Permanent versus determinant] \label{conj:permvdet}
The permanent of an $n \times n$ matrix $X$ cannot be written as the determinant of an $m \times m$ matrix $Y$ when the entries of $Y$ are linear combinations of the entries of $X$ and $m \leq n^{c}$ is polynomially bounded in $n$
\end{conjecture}

Mulmuley and Sohoni \cite{gct1} suggested a slight strengthening of this conjecture: roughly speaking, the permanent of an $n \times n$ matrix cannot even be approximated (in a certain sense made precise below) by a determinant of an $m \times m$ matrix under the conditions above.

To properly formulate this strengthened form of Conjecture~\ref{conj:permvdet}, we will now describe certain quasi-homogeneous algebraic varieties (i.e., orbit closures) which are the central objects of study in GCT. Let $V = \C^{m^{2}}$ be the vector space of $m \times m$ complex matrices. Let $x_{i,j}$ ($1 \leq i,j \leq m$) be a basis for its dual $V^{*}$. There is a naturally induced action of $\GL(V)=\GL_{m^{2}}(\C)$ on $\Sym^{m}(V^{*})$, the space of degree $m$ homogeneous polynomials in the $m^{2}$ variables $x_{i,j}$. The determinant polynomial $\det_{m}(X)$ is a point in this space, and we denote its orbit closure by $\Det_{m} := \overline{\GL(V) \cdot \det_{m}}$.

Since we wish to consider the permanent of a smaller $n \times n$ matrix, but $\perm_{n}$ is of lower degree than $\det_{m}$, we instead consider the ``blasted permanent'' $x_{m,m}^{m-n} \perm_{n}(X|_{n})$, essentially without loss of generality. Here $\perm_{n}(X|_{n}) := \sum_{\sigma \in S_{n}} x_{1, \sigma(1)} \dotsb x_{n, \sigma(n)}$, that is, we think of $X|_{n}$ as the upper-left $n \times n$ submatrix of an $m \times m$ matrix of variables. We denote the orbit closure of the blasted permanent by $\Perm_{n}^{m} := \overline{\GL(V) \cdot x_{m,m}^{m-n} \perm_{n}}$. Mulmuley and Sohoni's strengthening of Conjecture~\ref{conj:permvdet} is then:

\begin{conjecture}[Mulmuley and Sohoni \cite{gct1}] \label{conj:gct}
For each $c > 0$ and infinitely many $n$, $\Perm_{n}^{n^{c}} \not\subseteq \Det_{n^{c}}$.
\end{conjecture}

The geometry of a quasi-homogeneous variety can be studied via the representation theory of its coordinate ring and vice versa: since both $\Det_{m}$ and $\Perm_{n}^{m}$ contain dense open $\GL(V)$-orbits, their coordinate rings $\C[\Det_{m}]$, $\C[\Perm_{n}^{m}]$ are $\GL(V)$-modules. If $\Perm_{n}^{m} \subseteq \Det_{m}$, then $\mult(V_{\lambda}, \C[\Det_{m}]) \geq \mult(V_{\lambda}, \C[\Perm_{n}^{m}])$ for all irreducible representations $V_{\lambda}$ of $\GL(V)$. Mulmuley and Sohoni propose to separate these orbit closures by proving the existence of (for each $c > 0$ and infinitely many $n$) an irreducible $\GL(V)$-module whose multiplicity in $\C[\Perm_{n}^{m}]$ is strictly greater than its multiplicity in $\C[\Det_{m}]$ ($m=n^{c}$), thus violating the above condition and prohibiting an inclusion of varieties.

There are currently two important directions of research in the GCT program: (1) understand the representation theory of the coordinate rings of these orbit closures and (2) understand these orbit closures as geometric objects, e.g., by attempting to find defining equations, studying their differential structure and singularities, etc. These two directions are related, sometimes quite closely, but for the sake of presentation we have organized the rest of this section around these two themes.

\subsection{Representation Theory}
Before understanding the representations arising in $\C[\Det_{m}]$, a natural problem is to first understand the representations arising in the coordinate ring of the orbit $\GL(V) \cdot \det_{m}$, rather than the orbit closure. Sohoni, Weyman, Kumar, and Landsberg all touched on or focused on this topic in their presentations at the workshop.

The multiplicities $\mult(V_{\lambda}, \C[\GL(V) \cdot \det_{m}])$ are an upper bound for the multiplicities in the orbit closure, so even the multiplicities in the orbit of $\det_{m}$ could potentially be used to prove Conjecture~\ref{conj:gct}. The current computations and geometric results (discussed in the next section) suggest that the multiplicities in the orbit alone may not be a promising approach to Conjecture~\ref{conj:gct}, but this data could be reflecting misleading phenomena that only occur in very low degrees. Either way, understanding the representation theory of the orbit is an important first step in understanding the representation theory of the orbit closure.

By a standard result on algebraic group actions, $\C[\GL(V) \cdot \det_{m}] \cong \bigoplus_{\lambda} V_{\lambda} \otimes (V_{\lambda}^{*})^{G_{\det_{m}}}$, the sum of those $\GL(V)$-modules whose duals contain vectors invariant under $G_{\det_{m}}$, with each module appearing with multiplicity equal to the dimension of the space of $G_{\det_{m}}$-invariants. This stabilizer group was first calculated by Frobenius \cite{frobenius}: $G_{\det_{m}} = (\{(A,B) | \det(AB)=1\} / \langle (\alpha I, \alpha^{-1} I) \rangle \rtimes \Z/2\Z$, where $\Z/2\Z$ acts by transposition $X \mapsto X^{T}$, and the pair $(A,B)$ represents the map $X \mapsto AXB$. (Note that we had to take the quotient by $(\alpha I, \alpha^{-1}I)$ since this is the kernel of the action of $\GL_{m} \times \GL_{m}$ on $V^{*}$ by left and right multiplication $(A,B) \cdot X = AXB$.) The multiplicity of $V_{\lambda}$ in the coordinate ring of the orbit of $\det_{m}$ is thus the multiplicity of the trivial representation in the action of $G_{\det_{m}}$ on $V_{\lambda}$. By Schur--Weyl duality, the multiplicity of this action is the same as the following multiplicity for representations of the symmetric group:
\[
sk_{\lambda, (\delta^{m}), (\delta^{m})} := \mult_{S_{\delta m}}([\lambda], \Sym^{2}([(\delta^{m})]))
\]
where $[\lambda]$ denotes the irreducible representation of $S_{\delta m}$ corresponding to the partition $\lambda$, and $(\delta^{m})$ denotes the partition $(\delta, \delta, \dotsc, \delta)$. These multiplicities are referred to as ``symmetric Kronecker coefficients.'' 

If we ignore the action of transpose, the corresponding multiplicities for the symmetric groups are the Kronecker coefficients $k_{\lambda, \mu, \nu} := \mult_{S_{\delta m}}([\lambda], [\mu] \otimes [\nu])$ when $\mu = \nu = (\delta^{m})$. M. Sohoni discussed recent work \cite{sohoni} related to combinatorial understanding of the Kronecker coefficients. Using quantum groups, they have constructed crystal bases for the representations corresponding to Kronecker coefficients in which two of the partitions have at most two rows. This is in analogy with the case of Littlewood--Richardson coefficients, where such crystal bases have been constructed for all LR coefficients. The basic idea is that crystal bases translate questions about representations to purely combinatorial questions, which would hopefully be easier to understand. Sohoni also discussed some of the difficulties in extending these techniques to arbitrary Kronecker coefficients; see their paper \cite{sohoni} for details, as well as the related paper \cite{gct4}.

J.\ M.\ Landsberg mentioned that there is no known nontrivial $\lambda$ with $k_{\lambda, (\delta^{m}), (\delta^{m})} < \mult(V_{\lambda}, \C[\Sym^{m}(V^{*})]$, that is, all the Kronecker coefficients calculated so far theoretically or by computer cannot be used to resolve Conjecture~\ref{conj:gct}. However, with Ressayre (unpublished), they have found explicit nontrivial examples where the \emph{symmetric} Kronecker coefficients $sk_{\lambda, (\delta^{m}), (\delta^{m})}$ is less than the multiplicity of $V_{\lambda}$ in the whole space $\Sym^{m}(V^{*})$, and hence at least have a chance of being used to help resolve Conjecture~\ref{conj:gct}. Thus the experimental evidence so far suggests the importance of the action of the transpose when calculating multiplicities.

J.\ Weyman presented conjectures of Mulmuley on certain properties of the Kronecker coefficients. These conjectures also apply to sequences of multiplicities arising from the orbit closures of interest to GCT, but for simplicity we stick to the case of Kronecker coefficients here. The conjectures are modeled on results known to be true for the Littlewood--Richardson coefficients. The hope expressed by Mulmuley and Sohoni is that the (positive) resolution of these conjectures will eventually lead to the mathematics needed to resolve Conjecture~\ref{conj:gct}, though these conjectures do not directly imply Conjecture~\ref{conj:gct}.

To state the conjectures we need some preliminary definitions. A function $f\colon \N \to \Z$ is a \emph{quasipolynomial} if there are polynomials $f_{i}$ ($1 \leq i \leq \ell$) such that $f(n) = f_{i}(n)$ for all $n \equiv i \text{ mod } \ell$. A quasipolynomial is \emph{positive} if all the coefficients of each $f_{i}$ are nonnegative, and is $\emph{saturated}$ if $f_{i}(n) > 0$ whenever $f_{i}$ is not identically zero. Note that positivity implies saturation. The \emph{positivity index} $p(f)$ of a quasipolynomial $f$ is the least natural number $p$ such that $f(n + p)$ is positive, and the \emph{saturation index} $s(f)$ is defined similarly. 

For any fixed partitions $\lambda, \mu, \nu$, the function $k_{\lambda, \mu, \nu}(n) := k_{n\lambda, n\mu, n\nu}$ (where $n\lambda$ is the partition $(n\lambda_{1}, n\lambda_{2}, \dotsc)$) is known to be a quasipolynomial \cite{gct6}. 

\begin{conjecture}[Mulmuley \cite{gct6, BOR}] \label{conj:gct2}
Let $k(n) = k_{\lambda, \mu, \nu}(n)$. There are nonnegative constants $a,b$ (independent of $\lambda, \mu, \nu$) such that the saturation and 
positivity indices satisfy $s(k) \leq a h^b$ and $p(k) \leq a h^{b}$, where $h$ is the maximum height of $\lambda, \mu, \nu$. \end{conjecture}

The original conjecture \cite{gct6} was that $s(k) = 0$, but by analyzing the case where $\lambda, \mu$ have height two and $\nu$ has height three, Briand, Orellana and Rosas \cite{BOR} showed that the original conjecture was false, prompting Mulmuley to modify the conjecture to the form that appears here. The examples of Briand \emph{et al.} are counterexamples for the original conjecture, but not for Conjecture~\ref{conj:gct2}. Mulmuley also conjectured that the saturation index was zero for almost all partitions $\lambda, \mu, \nu$; see the appendix to \cite{BOR} for details.

Weyman mentioned that the cases analyzed by Briand--Orellana--Rosas fit into a more general framework, and it would be interesting to test simple cases of Mulmuley's conjectures in this more general framework, which we now discuss briefly. In particular, the Kronecker coefficients in which two partitions have two rows and the third partition has three rows are closely related to the orbits of $\GL_{2} \times \GL_{2} \times \GL_{3}$ acting in $\C^{2} \otimes \C^{2} \otimes \C^{3}$ in the obvious way. This is a particular example of a representation of a reductive group which has only finitely many orbits. 

Irreducible representations of reductive groups that have only finitely many orbits have been completely and beautifully classified, see e.\,g. \cite{kac,dadokKac} and sometimes go by the name ``Vinberg $\theta$ groups.'' Weyman suggests to first verify Mulmuley's conjecture for the orbit closures in modules with finitely many orbits, for several reasons. First, they are natural, beautiful examples that seem easier than the quasihomogeneous varieties needed for complexity lower bounds. In particular, the representations studied in GCT typically do not have finitely many orbits. Second, these examples are fairly well-understood. Third, because of the finiteness properties of these examples, verifying the analogue of Conjecture~\ref{conj:gct2} in these cases is a finite problem that can, at least in principle, be reduced to computer calculations. In practice, those calculations may turn out to be prohibitively large. Nonetheless, the Vinberg $\theta$-groups provide an interesting testing ground for Mulmuley's conjectures.

In a different direction, S.\ Kumar presented a new result \cite{kumar2} which shows that certain representations cannot be used to separate $\Det_{m}$ from $\Perm_{n}^{m}$ based solely on the presence or absence of these representations. Mulmuley and Sohoni conjectured not just that there are representations with $\mult(V_{\lambda}, \C[\Perm_{n}^{m}]) > \mult(V_{\lambda}, \C[\Det_{m}])$, but in fact that such representations exist with $\mult(V_{\lambda}, \C[\Det_{m}]) = 0$. We refer to such representations as ``incidence-based obstructions'' to the inclusion $\Perm_{n}^{m} \subseteq \Det_{m}$, since merely the presence or absence of the representation in the coordinate rings obstructs such an inclusion. Kumar showed that for any $\GL_{n^{2}}$ irreducible representation $V_{\lambda}$ with $\lambda = (\lambda_{1}, \lambda_{2}, \dotsc, \lambda_{n}, 0, 0, \dotsc, 0)$, $V_{n\lambda}$ appears in $\C[\Det_{m}]$, that is, its multiplicity is nonzero. This rules out a fairly large class of representations from being incidence-based obstructions, although it does not rule them out as being more general multiplicity-based obstructions.

Finally, K. Ye presented his work on the GCT of immanents \cite{ye}, which are a generalization of both the permanent and determinant. If $\chi$ is an irreducible character of the symmetric group $S_{n}$, then the immanent associated to $\chi$ is defined as $\imm_{\chi}(X) := \sum_{\sigma \in S_{n}} \chi(\sigma) x_{1,\sigma(1)} \dotsb x_{n, \sigma(n)}$. When $\chi$ is the trivial character, the immanent is the permanent and when $\chi$ is the sign character, the immanent is the determinant. Somewhere in between these extremes, the complexity of computing the immanents must change between $\cc{\# P}$-hard (as the permanent) and $\cc{P}$ (as the determinant). It is not known if ordering immanents by their computational complexity corresponds to any reasonable order in terms of characters, but some results of B\"{u}rgisser \cite{burgisser1,burgisser2} suggest that this is indeed the case (see also the recent result of Mertens and Moore \cite{mertensMoore}). Although it is suspected that separating immanents from one another is a more refined complexity question than permanent versus determinant (as their complexity is thought to lie between that of the determinant and permanent), immanents may still be fruitful to study from the viewpoint of GCT.


\subsection{Geometry}
We began the section on representation theory above by mentioning the representation theory of the orbit, rather than the orbit closure, of $\det_{m}$. Recall that the multiplicities in the orbit are an upper bound on the multiplicities in the orbit closure. It might be hoped that the multiplicities in the orbit, more than just an upper bound, might actually enable us to find directly the multiplicities in the orbit closure. One geometric property which has helped with exactly this problem in other situations is \emph{normality}.

Modules in the coordinate ring of the orbit are modules of regular functions -- that is, rational functions whose denominators do not vanish anywhere on the orbit. If every regular function in such a module extends from the orbit to the orbit closure -- that is, its denominator does not vanish on the boundary of the orbit -- then the module is also a submodule of the coordinate ring of the orbit closure. Determining which modules of regular functions extend from the orbit to the orbit closure is often referred to as ``the extension problem.'' 

Normality makes the extension problem easier to handle, as one consequence of normality is Hartog's Principle: any regular function defined on an open set extends to its closure, if the complement of the open set has codimension $2$ in its closure. In particular, in the case of $\Det_{m}$, this would imply that any regular function on the orbit which had no singularities in the orbits of codimension $1$ in the boundary would in fact extend to the entire orbit closure. 

However, S.\ Kumar has proven \cite{kumar} that none of the orbit closures $\Det_{m}$ ($m \geq 3$) and $\Perm_{n}^{m}$ ($n \geq 3$, $m \geq 2n$) are normal, putting the extension problem beyond currently standard techniques. This result exhibits the interplay between the representation theory and geometry well, both in that it involves the relationship between a geometric property (normality) and representations---the extension problem---and also in that the proof that these orbit closures is not normal uses representation theory.

K. Ye discussed some recent work which also exhibits the interplay between geometry and representation theory. He suggests that one might hope to separate $\Det_{m}$  from $\Perm_{n}^{m}$ by considering their tangent spaces and projective differential invariants. The tangent spaces are themselves modules for $G_{\det_{m}}$ (resp., $G_{\ell^{m-n}\perm_{n}}$), so these differential-geometric aspects of the quasihomogeneous varieties can be compared not merely in terms of their dimension, but also in terms of their representation-theoretic structure. Ye has begun determining the structure of the tangent spaces and projective differential invariants of these varieties.

J.\ M.\ Landsberg emphasized in his presentation the role of geometry in understanding the varieties of importance to GCT, suggesting that geometric insight should illuminate the representation theory, and that this philosophy has a role to play in GCT.

Despite the non-normality of the quasihomogeneous varieties studied by GCT, having a sufficiently good \emph{geometric} understanding of these varieties may still enable us to partially solve the extension problem, hopefully to a point that we would be able to resolve Conjecture~\ref{conj:gct}. There are (at least) two geometric tasks that are helpful for the extension problem: explicitly finding boundary components, and finding defining equations for these quasihomogeneous varieties.

The general technique for explicitly finding boundary components---as in \cite[Section~3.5]{LMR}, which is essentially the only known nontrivial example of a boundary component of $\Det_{m}$---is to find a (maximal) subspace of $V$ on which $\det_{m}$ vanishes, and picking a projection onto that subspace. Then by precomposing (the polarization of) $\det_{m}$ with this projection one gets a function in the boundary of $\Det_{m}$. Moreover all such functions arise this way. Understanding maximal subspaces of the zero locus of $\det_{m}$ is a centuries-old problem.

The basic idea to find defining equations of $\Det_{m}$ is to exploit properties specific to the determinant. In general, any unusual property or pathology of the determinant may lead to a defining equation for $\Det_{m}$. Landsberg and Ressayre have used determinantal identities such as Segre's identity $\det_{n^{2}}(\text{Hess}(\det_{n}(A))) = (-1)^{\binom{n+1}{2}} (n-1) (\det(A))^{n(n-2)}$ to find certain equations in the defining ideal of $\Det_{m}$. They have also discovered some new determinantal and \emph{permanental} identities, and have used these to get, among other equations, the first equations known to lie in the defining ideal of $\Perm_{n}^{n}$.

Landsberg also mentioned that Kumar's recent result (discussed in the previous section) can almost be recovered purely geometrically, assuming the Foulkes--Howe Conjecture \cite{FH}. 

Another interesting avenue of research mentioned briefly by Landsberg is the idea to leverage results on the asymptotic growths of multiplicities such as those by Brion \cite{brion}, Manivel \cite{manivel1, manivel2}, and Ressayre \cite{ressayre}, based on techniques by Dolgachev and Hu \cite{dolgachevHu}. Although exact multiplicities are often difficult to compute (both in practice in the sense of complexity theory), there are geometric techniques that sometimes allow one to estimate the asymptotic growth rates of certain multiplicities. Since GCT is ultimately concerned with asymptotic results in the geometry of an infinite family of varieties, these asymptotic, geometric techniques may have a role to play and the connection should be further explored.

\subsection{Open Problems}
Here we mention open problems related to GCT that arose during the presentations, discussions, and the open problem session at the workshop. In no particular order:

\begin{open}[S.\ Kumar]
The endomorphism orbit $\End(V) \cdot \det_{m}$ (the ``orbit'' under the action of all matrices, not just invertible matrices) may also be an affine algebraic variety, although it is not a closed subset of $\Sym^{m}(V^{*})$. For example, if $\Det_{m} \backslash \left(\End(V) \cdot \det_{m} \right)$, which is Zariski-closed, consists of codimension $1$ components whose union is carved out of $\Det_{m}$ by a single equation, then the endomorphism orbit would be an affine algebraic variety. If the endomorphism orbit were an affine algebraic variety, it would be useful to study this geometric object rather than the full orbit closure $\overline{\GL(V) \cdot \det_{m}} = \overline{\End(V) \cdot \det_{m}}$. In particular the extension problem here, from the $\GL$-orbit to the $\End$-orbit, is comparatively simpler. We remark that separating permanent from the endomorphism orbit of determinant is essentially a restatement of the original permanent versus determinant Conjecture~\ref{conj:permvdet} (unlike Conjecture~\ref{conj:gct}, which strengthens Conjecture~\ref{conj:permvdet}).
\end{open}

\begin{open}[J.\ Weyman]
Verify Mulmuley's conjectures (analogues of Conjecture~\ref{conj:gct2} above) for orbit closures in the representations of the Vinberg $\theta$-groups, that is, those with finitely many orbits. 
\end{open}

\begin{open}[P.\ B\"{u}rgisser, presented by J.\ M.\ Landsberg]
Study the geometric complexity theory of matrix multiplication. Consider the matrix multiplication tensor as a point in $M_{k \times \ell} \otimes M_{\ell \times q} \otimes M_{k \times q}^{*}$ and consider its orbit closure under the natural action of $\GL_{k\ell} \times \GL_{\ell q} \times \GL_{k q}$. This orbit closure should be compared against (equations for) the $r$-th secant variety of the Segre product of $\PP^{k \ell -1} \times \PP^{\ell q - 1} \times \PP^{kq-1}$, which consists of all points of border-rank at most $r$. See \cite{burgisserIkenmeyer} for details and some initial results.
\end{open}

\begin{open}[J.\ A.\ Grochow, presented by J.\ M.\ Landsberg]
Can we separate iterated $3 \times 3$ matrix multiplication from permanent or determinant? That is, consider the orbit closure of the function $(A_{1}, \dotsc, A_{n}) \mapsto \tr(A_{1} A_{2} \dotsb A_{n})$ where the $A_{i}$ are $3 \times 3$ matrices. This function is known to be complete for polynomial formula size \cite{benOrCleve}. The question of the formula size of permanent and determinant is classical and well-known in complexity theory; the suggestion here is that formula-size complexity, in the guise of the iterated $3 \times 3$ matrix multiplication, might be significantly amenable to the techniques of GCT.
\end{open}

\begin{open}[J.\ M.\ Landsberg]
What is the generic determinantal complexity of degree $d$ polynomials in $k$ variables? For example, $\Sym^{3}(\C^{4}) \subseteq \Det_{3}$; what is the smallest $n$ such that $\Sym^{d}(\C^{k}) \subseteq \Det_{n}$? This question is interesting even for small examples. In a similar vein, small examples of the determinantal complexity of the permanent are also potentially interesting: the blasted version of $\perm_{3}$ is known to lie in $\Det_{7}$ \cite{grenet} and to not lie in $\Det_{4}$ \cite{mignonRessayre}; what is in fact the smallest $n$ such that $z^{n-3}\perm_{3} \in \Det_{n}$?
\end{open}

\section{Positivity, Concavity, Complexity, and GCT}
L.\ Gurvits gave two talks around \emph{positivity} and \emph{(log-)concavity}; these two themes play an important role in the representation theory relevant to GCT, and also more directly in combinatorics and algorithms relating to graph matching problems (see, e.\,g. \cite{gurvitsHyp, gurvitsMatch}). Indeed, Mulmuley's positivity and saturation conjectures for multiplicities arising in GCT are partially motivated as a first step towards a positive combinatorial formula for these multiplicities. In other words, the multiplicities should count some type of combinatorial object. The hope is that such combinatorial understanding of the multiplicities would enable a direct combinatorial proof of the desired multiplicity inequalities for infinitely many $n$.

For example, in the case of Littlewood--Richardson coefficients, it is known that the LR coefficients---which are multiplicities in representation of $\GL_n$---in fact count the number of integer points in certain explicitly given polytopes \cite{BZ, knutsonTao}, related to the expression of LR coefficients as counting certain constrained Young tableaux. 

Okounkov \cite{okounkov} discusses why one might expect certain families of multiplicities to be log-concave, via an interesting connection with entropy and statistical physics. Indeed, Okounkov \cite{okounkov} showed that this is the case asymptotically. However, it was later discovered that this can not hold exactly \cite{CDW}. 

Gurvits \cite{gurvitsMatch} shows that a certain problem which is $\cc{NP}$-hard on general polynomials is in fact in $\cc{BPP}$ for strongly log-concave polynomials, and in fact is in $\cc{P}$ for $H$-stable polynomials. This is an example where log-concavity directly implies computational tractability, providing further evidence that log-concavity may be useful to find the efficient algorithms conjectured by Mulmuley and Sohoni. 

Three notions related to positivity and concavity arise in Gurvits's work: strong log-concavity, hyperbolicity, and $H$-stability. A polynomial is \emph{strongly log-concave} if all of its mixed partial derivatives of all orders are either identically zero, or log-concave on $\R^{n}_{\geq 0}$. A complex polynomial $p(z_{1}, \dotsc, z_{n})$ is \emph{hyperbolic in direction $(e_{1}, \dotsc, e_{n})$} if the univariate polynomial $p(z_{1} - t e_{1}, \dotsc, z_{n} - t e_{n})$ in the variable $t$ has only real roots whenever the $z_{i}$ are all real. If furthermore the roots in $t$ are all positive reals whenever the $z_{i}$ are positive, then $p$ is called \emph{$H$-stable}. $H$-stability implies strong log-concavity. Note that having all positive roots is a stronger condition than having all positive coefficients, which is a property Mulmuley conjectures for the stretching quasipolynomials related to GCT.

The computational problem which Gurvits shows tractable assuming log-concavity is: given a homogeneous polynomial $p(x_{1}, \dotsc, x_{n})$, is there a partition $\Gamma = (\Gamma_{1}, \dotsc, \Gamma_{k})$ of the variables such that $\sum_{i=1}^{k} \deg_{\Gamma_{k}}(p) = \deg(p)$? A simple reduction from 3-colorability shows that this problem is $\cc{NP}$-hard in general. On the other hand, when $p$ is strongly log-concave, Gurvits shows that this is equivalent to a problem known to be in $\cc{BPP}$, namely the separation of variables problem: given $p(x_{1}, \dotsc, x_{n})$ is there a partition $\Gamma$ as above such that $p$ factors as $p(x_{1}, \dots, x_{n}) = \prod_{i=1}^{k} p_{i}$ where, for all $i$, $p_i$ only involves variables in $\Gamma_i$? Such a factorization is possible if and only if the Hessian matrix of $p$ has a block-diagonal structure; this is completely determined by the $0$-pattern of the Hessian, which can be determined by a standard randomized algorithm. As mentioned above, if furthermore the polynomial $p$ is $H$-stable, then this algorithm can be derandomized. 

Gurvits also suggested that perhaps determinantal and permanental \linebreak \emph{in}equalities over $\R$ might be useful for proving complexity lower bounds over $\C$. He points out that, while the permanent has only a relatively small stabilizer, and very few known identities, it has many known semi-algebraic inequalities (that is, over $\R$). Moreover, it seems that permanental and determinantal inequalities have been more useful in combinatorics than have equalities. For example, the van der Waerden inequality states that the permanent of a doubly stochastic $n \times n$ matrix is at least $n! / n^n$. Gurvits \cite{gurvitsVDW} used strong log-concavity to generalize and simplify the proof of the van der Waerden inequality and also the Schrijver--Valiant Conjecture on the number of perfect matchings in $k$-regular bipartite graphs. 

Gurvits suggested that the Pascal determinant might be another polynomial that 
is fruitful to work with in the GCT setting. 
The Pascal determinant of a $(k+1)$-tensor $P$ is
\[
\pdet(P) := \sum_{\pi_{1}, \dotsc, \pi_{k} \in S_{n}} \text{sgn}(\pi_{1} \dotsb \pi_{k}) \prod_{i=1}^{n} P_{i, \pi_{1}(i), \dotsc, \pi_{k}(i)}
\]
Computing $\pdet$ is $\cc{\# P}$-hard.
Moreover, the Pascal determinant provides yet another link between positivity, polynomial identity testing for arbitrary polynomials over $\C$, and matching theory (not to mention quantum entanglement). 

When $P$ is an $n \times n \times n \times n$ $4$-tensor, we may think of $P$ as defining a map $M_{n} \to M_{n}$ as follows. Consider a ``flattening'' of $P$, as an $n^{2} \times n^{2}$ block matrix, which has $n \times n$ blocks each of size $n \times n$. Let $P^{(i,j)}$ denote the $n \times n$ block whose block-indices are $(i,j)$, that is, $P^{(i,j)}_{k\ell} = P_{ijk\ell}$. Then $P$ defines a map $M_{n}\to M_{n}$ by $X \mapsto T_{P}(X) := \sum_{i,j} X_{ij} P^{(i,j)}$. $P$ is Hermitian positive semi-definite if and only if $T_{P}$ is a so-called completely positive operator, i.e., $T_{P}(X)$ can be written $T_{P}(X) = \sum_{i=1}^{n} A_{i} X A_{i}^{*}$ for some complex matrices $A_{1}, \dotsc, A_{n}$. In this case, $\pdet(P) = || \det(\sum_{i=1}^{n} z_{i} A_{i}) ||_{F}$, where here the $z_{i}$ are independent variables, and $|| \cdot ||_{F}$ denotes the {\em Fischer norm} on polynomials. Namely, for a polynomial $\sum_{\omega_{1}, \dotsc, \omega_{n}} a_{\omega_{1}, \dotsc, \omega_{n}} x_{1}^{\omega_{1}} \dotsb x_{n}^{\omega_{n}}$, the Fischer norm squared is $\sum |a_{\omega_{1}, \dotsc, \omega_{n}}|^{2} \omega_{1}! \dotsb \omega_{n}!$. The Fischer norm is the unique unitarily invariant norm on the space of homogeneous polynomials of a fixed degree. Moreover, the Fischer norm of a polynomial is clearly zero if and only if the polynomial is zero, so testing whether $\pdet$ of a Hermitian PSD operator is zero is equivalent to the polynomial identity testing problem for the polynomial $\det(\sum z_{i} A_{i})$. This in turn is a reformulation of Edmonds's problem, which is to determine whether the span of the $A_{i}$ contains an invertible matrix. Thus $\pdet$ provides a link between positive operators and arbitrary complex matrices, which might be useful to prove complexity lower bounds over $\C$, given the $\cc{\# P}$-hardness of $\pdet$.

Using theory related to $\pdet$, Gurvits was able to derandomize an algorithm for a special case of Edmonds's problem, namely when the span of the $A_{i}$ has a basis consisting of rank $1$ matrices. We suspect that there are more and deeper connections here to be explored. See \cite{gurvitsMatch} for more details.

\section{The BSS model}
\newcommand{\cP}{\cc{P}}
\newcommand{\cNP}{\cc{NP}}
\newcommand{\ccoNP}{\cc{coNP}}

Another approach to the $\cc{P}$ vs.\ $\cc{NP}$ question is to look at the BSS
computational model. The idea is a simple generalization of the classical Turing
machine model. Start with a ring (or group), $R$, and, rather than restricting
the Turing tape to elements of $\{0,1\}$, one may put any element of $R$ onto the
Turing tape. If $R=\mathbb{F}_2$ then we obtain the classical Boolean
computational model as a special case. For other $R$ we may ask whether $\cc{P}_R$ and
$\cc{NP}_R$, or BSS analogues of other complexity classes, are equal.  Formally, $\cc{P}_R$ and $\cc{NP}_R$ can be defined as follows:

\begin{definition}
  Let $x\in R^n$ be our input and define $\size(x):=n$. For a given decision
  machine, $M$, denote $TM(x)$ as the number
  of arithmetic operations needed for $M$ to decide $x$.
  \begin{enumerate}
    \item
      $(X,X_{yes})\in\cc{P}_R$ if there is a decision machine, $M$, and a polynomial,
      $p$, such that $TM(x)\le p(\size(x))$.
    \item
      $(X,X_{yes})\in\cc{NP}_R$ if there exists $(Y,Y_{yes})\in\cc{P}_R$ and
      polynomial, $p$, such that $x\in X_{yes}$ implies there is a witness $w\in
      R^{p(\size(x))}$ with $(x,w)\in Y_{yes}$.
  \end{enumerate}
\end{definition}

Intuitively, $\cc{P}_R$ is the class of problems that can be decided in time
polynomial in the input size. We see $\cc{NP}_R$ as the class of problems whose
yes answers can be verified using a witness whose size is polynomial in the
input size; the verification also taking time polynomial in the input size.
It is clear that $\cc{P}_R\subseteq\cc{NP}_R$.
Given a class of machines $\cc{A}$ (e.g., as in the definition of $\cc{P}_{R}$) and a complexity class $\cc{B}$, we next define an oracle. The
complexity class $\cc{A}$ with an oracle $\cc{B}$, denoted $\cc{A}^{\cc{B}}$ is
the complexity class derived from a machine for $\cc{A}$ that can decide
problems in $\cc{B}$ in constant time. Thus $\cc{P}^{\cc{NP}}$ is the class of
decision problems that can decided in polynomial time assuming we can decide
$\cc{NP}$ problems for free.

\forkorben{{Lenore Blum}}
Lenore Blum discussed what one could call the fundamental $\cc{NP}$ problem in
the BSS model.  Let $f_1,\dots,f_m\in R[x_1,\dots,x_n]$ be a system of
equations. We want to know whether there is a solution, over $R$, to this system. This is
known as {\it Hilbert's Nullstellensatz} over $R$, or $\cc{HN}_R$. (This problem is sometimes referred to as {\em feasibility over $R$} or {\em Hilbert's Tenth Problem over $R$}.) It is easy to
see that a solution serves as an easily verifiable witness. Therefore
$\cc{HN}_R\subset\cc{NP}$. But we also have
\begin{theorem}[$\cNP$-completeness Theorem \cite{BSS89}]
  $\cHN_R$ is $\cNP$-complete over $R$ when $R$ is an integral domain, e.g., 
  $\BZ_2$, $\BR$, or $\BC$. \qed 
\end{theorem}

$\cc{HN}_R$ is decidable when $R=\BC$ and when
$R=\BR$, thanks to classical results of Hilbert and Tarski 
that ultimately formed the basis for computational algebra. On the other hand, 
$\cc{HN}_{\BZ}$ is undecidable, thanks to the famous 
results of Davis, Matiyasevich, Putnam, and Robinson on Diophantine 
equations \cite{matiyasevich70,DMR76}. Hence $\cc{NP}_\BZ\ne\cc{P}_\BZ$.

\forkorben{Mike Shub}

Another classical approach to the $\cc{P}_R$ vs.\ $\cc{NP}_R$ question was
presented by Mike Shub.  This is the 20 Questions problem. In this decision
problem one needs to decide whether an input complex number is a non-negative 
integer less than or equal to a given input integer, using only equality tests and field arithmetic---inequality tests involving $<$ or $>$ are not allowed. We also include a string of
ones of length $\lfloor\log(|k|+1)\rfloor$. The question is then whether we can
decide this question in polynomial time, hence polynomial in $\log(|k|+1)$. This decision problem is in
$\cc{NP}_\BC$.  Shub and Smale conjectured that it is not in $\cc{P}$.
\cite{SS95} He also presented the related $\tau$-conjecture, which implies
that 20 Questions is not in $\cc{P}$.

\begin{conjecture}[$\tau$-conjecture]\ind{$\tau$-conjecture}
  For any $f\!\in\!\BZ[T]$, we define $\tau(f)=\min\{k\; |\; f_k=f\}$ where 
$(f_{-1},f_0,\ldots,f_k)$ is any sequence satisfying 
$f_{-1}(T)=1$, $f_0(T)=T$, and for any $\ell,m<j\leq k$,  
we have $f_j=f_\ell\circ f_m$ where $\circ$ is $+$, $-$, or $*$. 
Finally, let $Z(f)$ be the number of {\bf distinct} integer zeros of
$f$. Then there is $c>0$ such that $Z(f)\le(1+\tau(f))^c$ for all $f$.
\end{conjecture}  

\noindent 
Shub concluded by discussing approaches toward finding 
polynomials with a number of integers roots exponential in $\tau(f)$. 

\forkorben{Thierry Zell}
\ind{$\cPH$}\ind{$\Sigma_i$}\ind{$\Pi_i$}
Recall that the polynomial hierarchy, $\cPH$, can be defined as follows: 
let $\cc{\Sigma}^{p}_0=\cNP$ and $\cc{\Pi}^{p}_0=\ccoNP$. We then inductively 
define $\cc{\Sigma}^{p}_i=\cNP^{\cc{\Pi}^{p}_{i-1}}$, 
$\cc{\Pi}^{p}_i=\ccoNP^{\cc{\Sigma}^{p}_{i-1}}$, and 
$\cPH=\bigcup_{i} \cc{\Sigma}^{p}_i = \bigcup_{i} \cc{\Pi}_{i}^{p}$.
Thierry Zell presented BSS analogues to Toda's theorem. Let $\#\cc{P}$ denote
the counting problem that counts the number of witnesses for a given decision
problem in $\cc{NP}$. Toda's theorem says that
$\cc{PH}\subset\cc{P}^{\#\cc{P}}$ \cite{toda91}. That is, the
ability to count is very powerful. For a BSS machine over an infinite ring the
number of solutions will most often either be 0 or infinite. This makes
$\#\cc{P}_R$ almost indistinguishable from $\cc{NP}_R$. For such $R$ 
it thus makes sense to ask about better-behaved geometric invariants 
(such as Betti numbers) instead of cardinalities of solutions sets. 
From this point of view, Zell and Basu
proved an analogue of Toda's theorem for compact subsets of
$\BR$ \cite{basuzell}. Later Basu proved the same theorem when $R$ is a compact
subset of $\BC$ (i.e., projective space or the unit ball) \cite{basucompact}. 

%

\forkorben{Peter Scheiblechner}
\newcommand{\BoolP}{\operatorname{BP}}

Peter Scheiblechner described the {\it Boolean part} of complexity classes in 
the BSS model, an idea coming from earlier work of Cucker and other authors. 
Here we allow arbitrary rings or groups inside the
machine as usual, but we restrict the inputs to 
$\bigcup_{n\in\mathbb{N}}\{0,1\}^n$. For a complexity class,
$\cc{A}$, we denote this class restriction by $\BoolP(\cc{A})$. This gives a tool
to transfer results between the BSS models and the classical model. For 
example, we have the following result. 
\begin{theorem}[Koiran 1993\cite{koiran93}] \mbox{} \\ 
\mbox{}\hfill $\BoolP(\cP_{(\BR,+,<)})=\cP$\mbox{}\hspace{.75cm}\mbox{} 
\hfill \mbox{}\\  
\mbox{}\hfill $\BoolP(\cNP_{(\BR,+,<)})=\cNP$ \hfill \qed 
\end{theorem}

Given a complexity class $\cc{A}$ over $\BR$, we can construct a
related class, the weak version of $\cc{A}$ denoted $\cc{A}_{W}$. This is 
the same class except we restrict repeated squarings. 
This reduces the maximum degree of the intermediate polynomials. We then 
have the following result:  
\begin{theorem}[Koiran 1993\cite{koiran93}] \mbox{}\\ 
\mbox{}\hfill $\BoolP(\cP_{W})=\cP/poly$\mbox{}\hspace{.3cm}\mbox{} 
\hfill \mbox{}\\  
\mbox{} \hfill $\BoolP(\cc{DNP}_W)=\cNP/poly$ \hfill \qed 
\end{theorem}

It is worth noting that $\cP_W$ is not equal to $\cNP_W$. On the other 
hand, showing $\cP/poly \neq \cNP/poly$ is a major open question, closely 
connected to $\cP \neq \cNP$.
\begin{theorem}[Cucker \cite{cucker}] \mbox{}\\ 
\mbox{}\hfill $\cP_W\ne\cc{DNP}_W\subset\cNP_W$ \hfill \qed 
\end{theorem}

For the other transfer results we need to define two decision problems. First 
$\cc{PosSLP}$ is the problem of deciding whether a straight line program
constructing an integer produces a positive integer. A related problem
$\cc{EquSLP}$ decides whether a straight line program produces $0$. Finally, we
define $\cc{P}_\BR^0$ to be the sub-class of $\cc{P}_{\BR}$ where only the constants $0$ and $1$ are allowed in computations. We then have the following result: 
\begin{theorem}[\cite{abkm09}]
  \begin{align*}
    \BoolP(\cc{P}^0_\BR)&=\cc{P}^{\cc{PosSLP}}\\
    \BoolP(\cc{P}_\BC)&=\cc{P}^{\cc{EquSLP}}\subset\cc{BPP}\\
    \BoolP(\cc{DNP}_\BC)&=\cc{NP}^{\cc{EquSLP}}\subset\cc{NP}^{\cc{coRP}}. \qed 
  \end{align*}
\end{theorem}

\subsection{Maurice Rojas: Number Theoretic Connections}
\forkorben{Maurice Rojas}
Maurice Rojas began by describing weakenings of the $\tau$-conjecture that 
may be more tractable or admit a broader set of tools. For instance, one 
can go farther and conjecture that the number of rational roots 
(as well as the number of integral roots) of $f$ is bounded above by a 
polynomial in $\tau(f)$. Going still farther, one can consider the number of 
roots of $f$ in a completion, such as $\mathbb{R}$ or the $p$-adic 
rationals $\mathbb{Q}_p$ for $p$ any prime. Unfortunately, such  
overly-optimistic conjectures are false: simple families of polynomials with 
a number of real roots exponential in $\tau(f)$ were found shortly 
after the original statement of the $\tau$-conjecture. Moreover, Poonen 
and Rojas later found examples of polynomials with a number of 
$p$-adic rational roots exponential in $\tau(f)$. However, such 
examples appear to have too many roots over only a small number of 
completions of $\mathbb{Q}$. 

With this in mind, Rojas stated an ``adelic'' form of the $\tau$-conjecture: 
\begin{conjecture}[adelic $\tau$-conjecture]\ind{adelic $\tau$-conjecture} 
There is an absolute constant $c$ such that,
for any $f\!\in\!\mathbb{Z}[T]$, there is a field
$K\!\in\!\{\mathbb{R},\mathbb{Q}_2,\mathbb{Q}_3,\mathbb{Q}_5,\dots\}$ 
such that $f$ has no more than $(\tau(f)+1)^c$ roots in $K$.
\end{conjecture}
\noindent Since $\BQ_p$ contains $\BZ$ then it is clear that the
adelic $\tau$-conjecture implies the classical $\tau$-conjecture.

Rojas then presented Boolean polynomial hierarchy containments based on the
generalized Riemann hypothesis (GRH) and a slightly weaker conjecture.  We begin
by defining $\FEAS_\BC$. Given any $F\in\bigcup_{k,n\ge1}(\BZ[x_1,\dots,x_n])^k$
decide whether $F$ has a root in $\BC$. This is called the {\it complex
feasibility problem} or $\FEAS_\BC$. This problem is slightly different from 
$\cc{HN}_\BC$ because $\cc{HN}_\BC$ is a question involving the BSS model 
over $\BC$, while $\FEAS_\BC$ is posed over the classical (Boolean) Turing 
machine model. Koiran proved that GRH implies $\FEAS_\BC\in \cc{AM}~
(\subseteq\cc{\Sigma}_2^p)$ \cite{koiranGRH}. This result is impressive 
because no containment of $\FEAS_\BC$ in the polynomial-hierarchy is 
currently known unconditionally. Assuming GRH, proving $\FEAS_\BC\notin\cP$ 
would then provide some evidence (through Boolean parts) that $\cP\ne\cNP$. 
Rojas also presented a conjecture on the density of primes 
(which could still hold even under certain failures of GRH) 
that would imply $\FEAS_\BC\in\cP^{\cNP^{\cNP}} 
(\subseteq \cc{\Sigma}_3^p \cap \cc{\Pi}_3^p)$. We state Rojas' conjecture 
in the Open Problems Section below. 

\subsection{Pascal Koiran: Transfer Theorems} 
Pascal Koiran then presented transfer results and connections between the 
Valiant and BSS models. First, we may view certain algorithms 
as decision trees. If 
we restrict only the {\it depth} of the tree, rather than the size, to be
polynomial in the input size then we have $\cc{VPAR}$ -- the parallel Valiant
complexity class. In the standard Valiant model the degree of the defining
varieties is polynomially bounded by $n$.  Another class can be derived from the
Valiant class, $\cc{VP}$, by removing this restriction on the degree of the
polynomials defining the varieties. This class is called $\cc{VP}_{nb}$. With
these two classes Koiran and Perifel \cite{kp06} found a transfer result between
the Valiant and BSS models. That is, $\cc{VP}_{nb}=\cc{VPAR}$ implies
$\cc{P}_\BR=\cc{NP}_\BR=\cc{PAR}_\BR$.
Koiran also looked at the $\cc{P}$ versus $\cc{NP}$ question in terms of
polynomial depth decision trees. The key question is: For a given $M$ does 
$\cc{NP}_M$ have
polynomial depth decision trees? If it does not then we have
$\cc{P}_M\ne\cc{NP}_M$. He discussed cases where we can separate $\cc{P}_M$ from
$\cc{NP}_M$ based on this question. This includes $M=(\BR,+,-,<)$.

%
%
\section{Saugata Basu: BSS Analogues of Valiant Classes}
Basu proposed analogues of Valiant classes for the BSS model. Considering BSS machines over $k$ 
with $k$ one of $\BZ/2\BZ,\BR$, or $\BC$, observe that such machines recognize a sequence of
subsets $(S_n\subseteq k^n)_{n\ge0}$. These sets $S_n$ have to be definable in 
the corresponding structure/language. For finite fields, they are finite sets. 
For the real closed field $\BR$ they are semi-algebraic 
sets.\ind{semi-algebraic sets} For the algebraically closed field $\BC$ they are constructible
sets.

\begin{proposition}
  Let $k=\BZ/2\BZ,\BR,$ or $\BC$.
  \begin{enumerate}
  \item
  There are $\cNP_k$-complete sets.
  \item
    The polynomial hierarchy, $\cPH_k$, is well-defined and $\cPH_k\subset\cc{EXP}_k$.
  \end{enumerate}
\end{proposition}

Note that when $\cPH_{k}\not\subseteq\cc{EXP}_{k}$ we get a separation $\cP_{k}\ne\cNP_{k}$. 
For example, we have $\cP_\BQ\ne\cNP_\BQ$ for $k=\BQ$, since $\mathbb{Q}$ does 
not admit quantifier elimination.

\subsection{Valiant's classes}
To simplify the discussion, for the time being, we will stick with $k=\BF_2$. An
element of $\cc{VP}$ for $\BF_2$ is a collection of functions
$(f_n:\{0,1\}^n\rightarrow\{0,1\})_{n\ge0}$ with efficient circuits.
Next we consider the class $\cc{VNP}$. We use projections to define it. We start
with a function 
\[f_{m+n}:k^m\times k^n\rightarrow R,\]
where $R$ is a ring. We then want a projection $g_n:k^n\rightarrow R$. We define
it via the push-forward of the projection. That is
\[g_n(x)=\int_{k^m}f_{m+n}(y,x)dy.\]
When $k=\BF_2$ we have $g_n(x)=\sum_{y\in\BF_2^m} f_{m+n}(y,x)$. Therefore, one
can set up a more general theory if on has
\begin{enumerate}
  \item a suitable class of functions and
  \item a notion of push-forward or projection via integration along fibers.
\end{enumerate}

The proposed definition for $\cc{VNP}$ would be the push-forward of $\cc{VP}$.

\subsection{Constructible Functions}
One needs a general class of functions that has
\begin{enumerate}
  \item a measure of complexity and
  \item a natural measure such that the function class is closed under
    integration.
\end{enumerate}
There is a standard such class in the sense of algebraic geometry. This is the
class of {\it constructible functions}.\ind{constructible functions}
\begin{definition}
  A semi-algebraically constructible function over a real closed field is
  $f:R^n\rightarrow R$ that is a finite linear combination of characteristic
  functions $1_X$ of semi-algebraic sets $X\subseteq R^n$, e.g.
  $f=\sum_{i=1}^r a_i1_{X_i}$.
\end{definition}
\begin{remark}
  Other spaces have analogous constructions.
  \begin{enumerate}
    \item We have constructible sets for algebraically closed fields.
    \item For arbitrary O-minimal structures in model theory there are
      definable sets.
  \end{enumerate}
\end{remark}

Constructible functions form an $R$-algebra.
\begin{example}\mbox{}\\ 
  \begin{enumerate}
    \item If $X\subseteq R^n$ is semi-algebraic then $1_X$ is constructible.
    \item Sums and products of constructible functions are constructible.
    \item They form an infinite dimensional $R$-algebra generated by $\{1_X\}$.
    \item Let $P_{d,n}$ be the vector space of polynomials in $R[x_1,\dots,x_n]$
      of degree at most $d$. Let $V_R(p)=\{x\in R|f(x)=0\}$. We have the 
      following constructible functions
      \[f:P_{d,1}\rightarrow R,p\mapsto\#V_R(p).\]
      For arbitrary $n$ we have
      \[f:P_{d,n}\rightarrow R,p\mapsto\#\mathrm{connected~components~ of}~V_R(p).\]
      or more generally, $p\mapsto b_i(V_R(p))$, the $i^{th}$ Betti number of
      $V_R(p)$.
  \end{enumerate}
\end{example}

\subsection{Euler--Poincar\'e Characteristic/Measure}
A key fact (see, for example, \cite[p. 400]{Kashiwara-Schapira}) is that constructible functions can be integrated against 
the (generalized) Euler--Poincar\'e characteristic.

\begin{definition}
  \ind{Euler-Poincar\'e Characteristic} A generalized Euler-Poincar\'e
  characteristic
  \[\chi:\{\text{semi-algebraic sets}\}\rightarrow\BZ\] has the properties
  \begin{description}
    \item[Unit Interval] $\chi([0,1])=1$.
    \item[Strict Additivity] $\chi(A\cup B)=\chi(A)+\chi(B)-\chi(A\cap B)$.
    \item[Multiplicativity] $\chi(A\times B)=\chi(A)\chi(B)$.
  \end{description}
\end{definition}

\begin{proposition}
  There exists a generalized Euler-Poincar\'e characteristic, $\chi$, and it is
  unique.
\end{proposition}

\begin{proposition}
  The generalized Euler-Poincar\'e characteristic is invariant under
  semi-algebraic homomorphism.
\end{proposition}
{\bf Warning:} The generalized Euler-Poincar\'e characteristic is {\em not} homotopy invariant.

\begin{example}\mbox{}\\  
  \begin{enumerate}
    \item
      $\chi(*)=\chi(*)\chi([0,1])=\chi(*\times[0,1])=\chi([0,1])=1$
    \item
      $\chi([0,1]^n)=\chi([0,1])^n=1$
    \item
      $\chi( (0,1))=\chi([0,1])-\chi(\{0\})-\chi(\{1\})=-1$
    \item
      $\chi(\BR)=\chi( (0,1))=-1$
    \item
      $\chi(\BR^n)=\chi(\BR)^n=(-1)^n$
    \item
      \begin{align*}
        \chi(S^n)&=2(\chi(\bar{B}^n))-\chi(S^{n-1})\\
        &=2-\chi(S^{n-1})\\
        &=\left\{
        \begin{array}{cl}
          0&n~is~odd\\
          2&n~is~even.
        \end{array}
        \right.
      \end{align*}
  \end{enumerate}
\end{example}

\begin{definition}\ind{constructible function}
  Let $f:R^n\rightarrow R$ be constructible. Then $f$ can be written
  (non-uniquely)
  \[f=\sum a_i1_{X_i}.\]
\end{definition}

\begin{proposition}
  We can define $\int fd\chi=\sum a_i\chi(X_i)$. This is well-defined and
  independent of the expression chosen for $f$.
\end{proposition}

The following is a classical result.
\begin{proposition}
  Let $\pi:R^m\times R^n\rightarrow R^n$ be a projection and let $f:R^m\times
  R^n\rightarrow R$ be constructible. Then $g(x)=\int_{R^m}f(y,x)d\chi_y$ is a
  constructible function.
\end{proposition}

Next we discuss notions of complexity and size in this framework. It is clear
that the characteristic functions $\{1_{p_i\ge0}\}$ generate the $R$-algebra of
constructible functions. Hence we can write $f=\sum a_i1_{p_i\ge0}$. Then we
define size to be $\size(1_{p\ge0})=\size(p)$ in the sense of circuit
complexity. We also want to require the following constraints
\begin{itemize}
  \item $\size(c\varphi)\le\size(\varphi)$ for $c\in R$
  \item $\size(\varphi\psi)\le\size(\phi)+\size(\psi)$
  \item $\size(\phi+\psi)\le\size(\phi)+\size(\psi)$
\end{itemize}
For arbitrary constructible $f$, define $\size(f)$ to be the infimum of
possible decompositions of $f$.

Finally we are prepared to define analogues for the Valiant complexity classes.
\begin{definition}
  $\cc{VP}_R^*$ is the collection of families $(f_n:R^n\rightarrow R)_{n\ge0}$ of
  constructible functions with $f_n$ bounded in size by $\poly(n)$. Then
  $\cc{VNP}_R^*$ is the push-forward of $\cc{VP}_R^*$ under $\pi$.
\end{definition}

\begin{problem}[Open]
  Is $\cc{VP}_R^*\subsetneq\cc{VNP}_R^*$?
\end{problem}

\subsection{Towards a Sheaf-Theoretic Definition}
We want to extend this definition to constructible sheaves. The details of a
measure of complexity have not fully been worked out in this context. One needs
to go to the (bounded) derived category of sheaves.

\newcommand{\Ob}{\operatorname{Ob}}
\begin{definition}
  Take $X=\BR^n$ and a ring $A=\BQ$. Let $\Ob(D^bX)$ be the collection of
  bounded complexes of $A$-modules. Then $F\in\Ob(D^bX)$ is {\it constructible}
  if there is a semi-algebraic stratification $X=\cup X_i$ such that the stalks
  of the sheaves are constant on strata, and if the the cohomology of the
  stalks of $F$ are finitely generated. (The latter requirement is to ensure
  a well-defined for the Euler-Poincar\'e characteristic.)
\end{definition}

There is then an analogue of the Tarski--Seidenberg theorem.
\begin{theorem}
  There is a commutative diagram
  \[
  \xymatrix{
  \Ob(D^b(Y))\ar[r]^{R_\pi}\ar[d]&\Ob(D^b(X))\ar[d]\\
  \CF(Y)\ar[r]^{\int f(\cdot,y)d\chi}&\CF(X)
  }
  \]
  where $\CF(X)$ is the algebra of constructible functions on $X$ and $R_\pi$ is
  the direct image of the projection, $\pi$.
\end{theorem}

From here one only needs to formulate a notion of complexity in this more
general context.

\section{Open Problems}
\begin{open}[J. Maurice Rojas]
  Given any finite algebraic extension $K$ of $\mathbb{Q}$, define
  \[m_K(c)=[\log(3n_K\log d_K)]^c,\] 
  where $n_K$ and $d_K$ respectively denote the field extension  
  degree and the discriminant of $K$. 
  Then there are absolute constants $c,\delta>0$ such that for all such 
  $K$, and any $N\ge 2^{m_K(c)}$, at least $\delta N$ of the 
  intervals $[1+x^{m_K(c)},(1+x)^{m_K(c)}]$ (with  
  $x\in\{1,\ldots,N\}$) contain the norms of at
  least $1+n_K(x+1)^{m_K(c)/2}$ prime ideals in the ring of algebraic integers 
  $\mathcal{O}_K$. This conjecture
  would imply that $\FEAS_\BC\in\cc{P}^{\cc{NP}^\cc{NP}}$. While this 
  conjecture currently appears to be out of reach, it can still hold under 
  various failures of GRH. 
\end{open}

\begin{open}[Mike Shub]
  Let $\sigma(f)$ denote the additive complexity of $f\in\BZ[x]$. Then
  $\sigma(f)\le\tau(f)$. Do there exist sequences of polynomials exhibiting
  a number of integer roots exponential in $\sigma(f)$? 
\end{open}

\begin{open}[Pascal Koiran]
  Let $f(x),g(x)\in\BR[x]$, each with at most $T$ monomials. What is the best
  upper bound on the number of $\BR$-roots of $fg-1$? The best known bound is
  $2 T^{2} + 1$. Furthermore, is $O(T)$ a possible bound? An important motivation is 
  recent work of Koiran and co-authors showing that sufficiently sharp upper 
  bounds for the number of real roots of polynomials defined by certain 
  arithmetic circuits imply major new separation results in the direction of 
  $\cc{VP}\neq\cc{VNP}$.  
\end{open}

\begin{open}[Peter Scheiblechner]
  Under GRH, it is known that $\cc{HN}_\BC\in\cc{AM}$ \cite{koiranGRH}. What
  about $\forall\cc{HN}_\BC$? The problem $\forall\cc{HN}_\BC$ is deciding the truth of sentences of the form
  \[(\forall x\in\BC^m\exists y\in\BC^m~\bigwedge_i f_i(x,y)=0.\]
  It is known that $\forall\cc{HN}_\BC\in\cc{PSPACE}$. But is it in $\cc{PH}$
  (assuming GRH, if need be)?
  The approach of Koiran seems not to work here, because $\forall x\exists y$
  such that $y^2=x$ is true over $\BC$, but false for every $\BF_p$.
\end{open}

\begin{open}[Peter Scheiblechner]
  Another variation of $\cc{HN}_\BC$ is $\cc{UHN}_\BC$. This problem is: given a system of polynomial equations with integer coefficients with the promise that the system has at most one solution, does it have a solution? Is $\cc{UHN}_\BC\in\cc{AM}$? What about $\cc{UHN}_k$ for a field of low degree? A fact which suggests why $\cc{UHN}_\BC$ might be in $\cc{AM}$ unconditionally, e.\,g. without GRH, is that if a unique solution exists it is necessarily rational. This also suggests the more general question wherein the promise is a bound on the degree over the rationals of any possible solution.
\end{open}

\begin{open}[J. Maurice Rojas]
  What Galois groups, $G$, for the extensions of $\BQ$ derived from the coordinates of isolated solutions of polynomial systems, allow the corresponding restriction of 
  $\cc{HN}_\BC$ to be solved
  unconditionally? For example, consider the following problem, which we call $\cc{CYCLO}$: 
  if $F\in \bigcup_{k,n\in \mathbb{N}} (\BZ[x_1,\cdots,x_n])^k$ does the system
  have a solution with each $x_i$ a root of unity? (The resulting 
  Galois groups are cyclic.) Cheng, Tarasov, and 
  Vyalyi have since proved that this problem is in $\cc{NP}$ 
  \cite{CTV10}. But what about, say, bicyclic Galois groups? 
\end{open}

\begin{open}[Saugata Basu]
  Let $\FEAS_4$ be the feasibility problem of degree 4 polynomials over
  $\BR^n$. This problem is $\cc{NP}_\BR$-complete. Of course, $\cc{HN}_\BC$ is
  $\cc{NP}_\BC$-complete.  It is not known whether the compact versions of any
  problems are complete (including these problems). That is, where one checks
  for solutions in a given compact subset. Can one find a compact problem that
  is complete for either $\cc{NP}_\BR$ or $\cc{NP}_\BC$?  Similarly, can one
  prove whether the compact quadratic feasibility is complete for
  $\cc{NP}^c_\BR$ or whether the compact version of $\cc{HN}_\BC$ is complete
  for $\cc{NP}^c_\BC$?
\end{open}

The following problems were not stated during the talks but were 
posed during several lunches: 
\begin{open}[Pascal Koiran]  
Can one more easily prove upper bounds, in the spirit of Open Problem 
8, over the $p$-adic rationals? How about just for the number of 
distinct valuations of the roots? 
\end{open}  

\begin{open}[Pascal Koiran] 
A variant of the classical subset sum problem asks: given real (or complex) 
numbers $x_1,\ldots,x_n$, is there a subset with sum $1$? (The classical 
version instead has integral $x_i$ and asks for a subset with sum an 
input integer.) The known lower and upper bounds for the classical 
version are respectively $\Omega(n^2)$ and $O(n^4)$, in the computation 
tree model. Can one obtain a 
better lower bound for the real or complex variant, over the BSS model? 
\end{open} 

To state our final open question, first observe that, for any prime $p$, any 
function $f : \{0,1\}^n \longrightarrow \mathbb{F}_p$ can be represented
uniquely as a polynomial in $\mathbb{F}_p[x_1,\ldots,x_n]/\langle x^2_1-x_1,
\ldots,x^2_n-x_n\rangle$. So we can naturally define $\deg f$ as the
degree of the unique such polynomial representing $f$. For instance,
the degree of the $n$-ary AND function is $n$ since AND can be represented
by $x_1\cdots x_n$. It is not much harder to show that the PARITY function 
has degree $n$ if $p$ is odd. 

Now, an even more flexible method of representing such functions $f$ makes 
use of a subset of $A\subseteq \mathbb{F}_p$, in addition to a polynomial 
$\mathcal{P}\in \mathbb{F}_p[x_1,\ldots,x_n]/\langle x^2_1-x_1,
\ldots,x^2_n-x_n\rangle$. In particular, we say that the pair 
$(\mathcal{P},A)$ represents $f$ when $f(x)=1$ if and only if 
$\mathcal{P}(x)\in A$. We then define $\deg^A_p(f)$ to be the 
minimum of $\deg \mathcal{P}$ over all pairs $(\mathcal{P},A)$ 
representing $f$. As an example, it is not hard to show that 
$\deg^A_p(AND)\geq\frac{n}{p-1}$ for all $A$, using Fermat's Little Theorem. See Regan's survey \cite{regan} for a more extensive discussion of the uses of such representations and their generalizations in complexity theory. 

\begin{open}[Arkadev Chattopadhyay]   
If we make the obvious extension of $\deg^A_p(f)$ to non-prime 
moduli, what is $deg^A_m(AND)$ when $m$ is a product of $t$ distinct 
primes? Currently, it is known that there exist $A$ for which 
$deg^A_m(AND)=O\!\left(n^{1/t}\right)$. Also, it is known that 
for all $A$, $deg^A_m(AND)=\Omega\!\left(\log^{1/(t-1)}n\right)$. 
Can we bridge the gap any further? 
\end{open}  

Chattopadhyay adds that these questions are stepping stones towards understanding various constant-depth circuit classes. Indeed, it is still unknown 
whether NP can be separated from certain circuit complexity classes 
that appear to have extremely low complexity. Determining which Boolean 
functions have fibers that can be expressed as disjoint unions of fibers of 
``simple'' polynomials over $\BZ$ is one way to approach this problem.  

\section*{Acknowledgments}
We would like to thank Jeff Lagarias for his copious, detailed notes---we consider him a co-scribe-at-large for the workshop. We would also like to thank all of the workshop participants for such interesting presentations and conversations. We would like to thank the workshop organizers---J.\ M.\ Landsberg, 
J.\ Maurice Rojas, and Saugata Basu---at the very least for organizing the 
workshop, for selecting us as scribes, and for improving the presentation of 
these notes. Finally, we would like to thank ICERM for hosting and supporting the workshop and these notes.

\bibliographystyle{ams-alpha}
\bibliography{icerm} 
\end{document}